\documentclass[letter,twocolumn]{jpsj2} 

\usepackage{graphicx}

\title{Interaction and Localization of
One-electron Orbitals in an Organic Molecule:\\
Fictitious Parameter Analysis for Multi-physics Simulations}

\author{
Toshiya \textsc{Takami}\thanks{E-mail address: takami@cc.kyushu-u.ac.jp},
Jun \textsc{Maki}\thanks{E-mail address: maki-j@cc.kyushu-u.ac.jp},
Jun-ichi \textsc{Ooba},
Taizo \textsc{Kobayashi},
Rie \textsc{Nogita},
and Mutsumi \textsc{Aoyagi}}

\inst{Computing and Communications Center, Kyushu University,
Hakozaki 6--10--1, Higashi-ku, Fukuoka 812-8581, Japan}

\abst{We present a new methodology to analyze complicated
multi-physics simulations by introducing a fictitious parameter.
Using the method,
we study quantum mechanical aspects of an organic molecule in water.
The simulation is variationally constructed from
the {\it ab initio\/} molecular orbital method
and the classical statistical mechanics
with the fictitious parameter
representing the coupling strength between solute and solvent.
We obtain a number of one-electron orbital energies
of the solute molecule derived from the Hartree-Fock approximation,
and eigenvalue-statistical analysis developed in the study
of nonintegrable systems is applied to them.
Based on the results, we analyze localization properties
of the electronic wavefunctions under the influence of the solvent.
}

\kword{eigenvalue statistics, avoided crossings, quantum chaos,
multi-physics simulation, variational construction of coupled simulation,
electronic states of bio-molecules, one-electron orbital energy,
solute-solvent system, RISM/SCF}

\begin{document}
\maketitle

The multi-physics simulation is one of
the powerful methods to construct complex simulations
representing realistic systems.
Such a calculation is often constructed
by combining multiple theories with different scales of description
based on different approximations,
e.g., climate simulations by fluid dynamics of the atmosphere
surrounded by various external heat and humidity sources\cite{AGCM},
quantum materials bound
to classical large degrees of freedom\cite{Ogata,Sato}, etc.
Since reality and accuracy are required,
those simulations have been larger and more complicated year by year.
Then, guaranteeing their convergence and reliability
becomes a more difficult problem.

In this letter, we study electronic states of a peptide in water
obtained by a multi-physics simulation which consists of
{\it ab initio\/} molecular orbital (MO) methods
and classical statistical mechanics.
The quantum chemical nature of large molecules in a solvent
is also one of the most popular topics in physics and chemistry.
In the present calculation,
the simulation is constructed from a combination of
the self-consistent field (SCF) calculation
under the Hartree-Fock (HF) approximation\cite{SCF}
and the statistical mechanics calculation by three dimensional version
of the reference interaction site model (3D-RISM)\cite{rism}.
Although the coupled simulation of RISM and SCF itself
was developed in 1990's\cite{Ten-no}
and has been applied to several molecular systems,
it attracts public attention again due to the recent developments
of the distributed computing environments\cite{grid}.
Since the 3D-RISM/SCF coupled simulation\cite{Sato-Kovalenko}
heterogeneously couples
the different theories both of which require large computational resources,
the execution of the simulation for large solute molecules
is difficult not only in an arrangement of the computer resources
but also in theoretical aspects of computation
such as stability of the convergence and reliability of the simulation.

The aim of this letter is two fold:
one is to establish the methodology to ensure a proper execution
of the complicated simulation,
and the other is to introduce the new analysis
for a electronic state of molecules using eigenvalue statistics.\cite{Haake}
The target of our analysis is one-electron orbital
energies $\varepsilon_j$ and its wavefunction $|\phi_j\rangle$
obtained as spatially independent wavefunctions
under HF approximation.
Although $\varepsilon_j$ and $|\phi_j\rangle$ are
introduced through the variational minimization of the total energy,
they still retain physical reality without Koopman's theorem
if we consider energies and wavefunctions of
quasiparticles and quasiholes\cite{HF-statistics}.
Our key to accomplish these aims
is an introduction of a fictitious parameter
which can be used not only to investigate the stability
and continuity of the results
but also to perform sophisticated analyses
\cite{GRMN1990,TTcurvature,TTcharacter,ZD93c,ZD93a,MS2005}
based on the eigenvalue statistics.

At first, we construct 3D-RISM/SCF coupled simulation
according to the standard variational way used in 1D-RISM/MCSCF
\cite{sato-hirata-kato}.
In the present calculation,
the electronic states of the peptide are obtained
by SCF\cite{SCF} under the restricted HF approximation (RHF),
i.e., all electrons are assumed to configure closed orbitals.
The variational functional is chosen as
a total Helmholtz free energy\cite{rism,sato-hirata-kato}
\begin{equation}
\label{eqn:free-energy}
  A(\lambda)=E_{\rm solute}(\{|\phi_j\rangle\})
   +\Delta\mu(\{|\phi_j\rangle\}; \lambda),
\end{equation}
where $E_{\rm solute}$ is an energy of the solute molecule
and $\Delta\mu$ is the excess chemical potential
of the solute obtained in 3D-RISM with the coupling strength $\lambda$.
Statistical correlation functions of the solvent
are calculated under the charge distribution of the solute
represented by a set of the orbitals $\{|\phi_j\rangle\}$.
The variation of the first term in {\it r.h.s\/} of (\ref{eqn:free-energy})
with respect to $|\phi_j\rangle$ gives usual SCF equations for the MO theory
while the solvation effect is introduced through the second term.
Variation with constraints using Lagrange multipliers
gives conditions to be satisfied.
Thus, the total free energy (\ref{eqn:free-energy})
of the coupled system is stabilized by solving these conditions.

One of the merits of the variational construction of complicated simulations
is that a set of equations to be solved is given by a simple calculus
once we define a functional of the whole system.
In the present case, we only have to define
the solute-solvent interaction potential
\begin{equation}
  u_\gamma({\bf r};\lambda)=\lambda\sum_a\left[
    \frac{q_aq_\gamma}{r_a}
    +4\epsilon_{a\gamma}\left\{
       \left(\frac{\sigma_{a\gamma}}{r_a}\right)^{12}
      -\left(\frac{\sigma_{a\gamma}}{r_a}\right)^{6}
    \right\}
  \right]
\end{equation}
where $q_a$ is a point charge on a solute atom $a$,
$q_\gamma$ is a partial charge on a solvent site $\gamma$,
$r_a\equiv|{\bf R}_a-{\bf r}|$ is a distance between
the solute atom $a$ and the solvent site $\gamma$,
$\epsilon_{a\gamma}$ and $\sigma_{a\gamma}$ is given
by the standard parameter sets of Lennard-Jones potentials.
We adopt the method of Mulliken population
among several schemes to define $q_a$,
\begin{equation}
  q_a=Z_a-\sum_{j=1}^{n/2}\sum_{\nu\in a}\sum_{\nu'}\left[
    C^*_{\nu j}C_{\nu'j}\langle\chi_\nu|\chi_{\nu'}\rangle+\hbox{c.c.}
  \right]
\end{equation}
where $C_{\nu j}$ are coefficients in a basis-set expansion of $|\phi_j\rangle$
\begin{equation}
\label{eqn:basis}
  |\phi_j\rangle=\sum_\nu|\chi_\nu\rangle C_{\nu j}
\end{equation}
by the atomic basis function $|\chi_\nu\rangle$.
This representation reduces computational costs
of the whole simulation,
and can be used even in large-scale simulations for proteins.
Although the reduction of the cost
is often a result of compensation in accuracy,
it can be retained by combining other schemes of
the potential representation \cite{calc-3drism}.

After the variation with respect to $C_{\nu j}$,
the solvated Fock matrix element $F_{\mu\nu}$ between $\mu$ and $\nu$
is given by
\begin{equation}
\label{eqn:fock}
  F_{\mu\nu}(\lambda)=F^{(0)}_{\mu\nu}-\lambda V_{\mu\nu}S_{\mu\nu},
\end{equation}
where the overlap integral $S_{\mu\nu}$ is defined by
$\langle\chi_\mu|\chi_\nu\rangle$.
$V_{\mu\nu}$ represents the environmental potential term
induced by the partial charge distribution of water molecules,
\begin{equation}
  V_{\mu\nu}
   =\frac{1}{2}\int\left[
    \frac{1}{|{\bf R}_\mu-{\bf r}|}+\frac{1}{|{\bf R}_\nu-{\bf r}|}
  \right]Q({\bf r};\lambda)\ d{\bf r},
\end{equation}
where ${\bf R}_\mu$ and ${\bf R}_\nu$ represent atomic centers
of the basis functions $|\chi_\mu\rangle$ and $|\chi_\nu\rangle$,
respectively, and
\begin{equation}
  Q({\bf r};\lambda)
  \equiv\sum_\gamma\rho_\gamma q_\gamma g_\gamma({\bf r};\lambda)
\end{equation}
is the partial charge distribution
calculated by summing partial charges given by the distribution
function $g_\gamma({\bf r};\lambda)$ over all the solvent sites $\gamma$.
We can numerically solve the Roothaan equation\cite{SCF}
\begin{equation}
\label{eqn:roothaan-hall}
  {\bf F}(\lambda){\bf C}={\bf S}{\bf C}\varepsilon
\end{equation}
to obtain coefficients $C_{\mu\nu}$ of the basis-set expansion
as elements of the matrix ${\bf C}$ and the orbital
energies $\varepsilon_j$ as diagonal elements of $\varepsilon$.

\begin{figure}[tb]
\begin{center}
\includegraphics[scale=0.5]{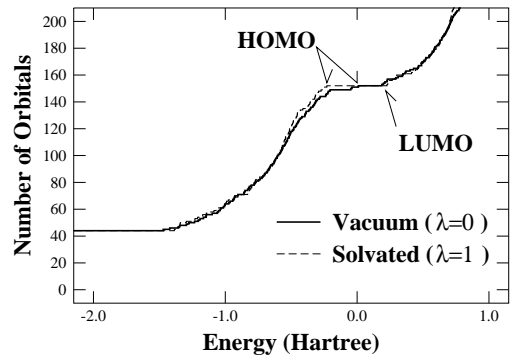}
\caption{The integrated density of one-electron orbitals
of a met-enkephalin. Two curves (vacuum and solvated) are plotted
around the area near the orbitals of HOMO and LUMO.}
\label{fig:staircase}
\end{center}
\end{figure}

The actual procedure to execute the coupled simulation is the following.
The SCF calculation is performed from an appropriate
initial distribution of the solvent.
For the given distribution of partial charges of the solvent,
the electron density is obtained after a convergence of the SCF.
Then, 3D-RISM calculation to obtain $g_\gamma({\bf r})$
is carried out using the charge distribution of the solute molecule.
Converged $g_\gamma({\bf r})$ determines
the environmental term $\lambda V_{\mu\nu}$ in the solvated-Fock matrix
by which the SCF calculation is performed again.
This procedure repeats until the whole distribution is converged.
From the converged simulation,
we obtain the solute electronic states
and the solvent distribution functions.

Before entering the analysis using $\lambda$,
we show results of a methionine-enkephalin in two specific cases.
This peptide consists of a sequence of 5 residues,
Tyrosine, Glycine, Glycine, Phenylalanine, and Methionine,
with 75 atoms in total.
From the standard SCF calculation, which corresponds to $\lambda=0$,
the orbital energies of the methionine-enkephalin are obtained in vacuum,
while the fully-solvated case $\lambda=1$ gives
the orbital energies in aqueous solution.
For the electronic ground state, we assume that
304 electrons occupy 152 independent spatial orbitals under RHF.
Among several band-like structures in the orbital energies,
we concentrate the analysis of the band just below the
highest occupied molecular orbital (HOMO) and the lowest
unoccupied molecular orbital (LUMO),
which contains 108 orbitals from 45th to 152th ones.
Orbitals strongly localized on 1s atomic orbital of C, N, O,
and 1s, 2s, 2p orbitals of S (known as core orbitals)
can be found in the region far from the band we analyze.
Since they exhibit almost no interaction with other orbitals,
we do not consider these orbitals in the present analyses.
The integrated density of states, $D(E)=\sum\Theta(E-E_j)$,
is shown for the band around HOMO and LUMO in Fig.~\ref{fig:staircase}.
In this case, the 152nd orbital is HOMO and the 153rd is LUMO.

\begin{figure*}[tb]
\begin{center}
\includegraphics[scale=0.5]{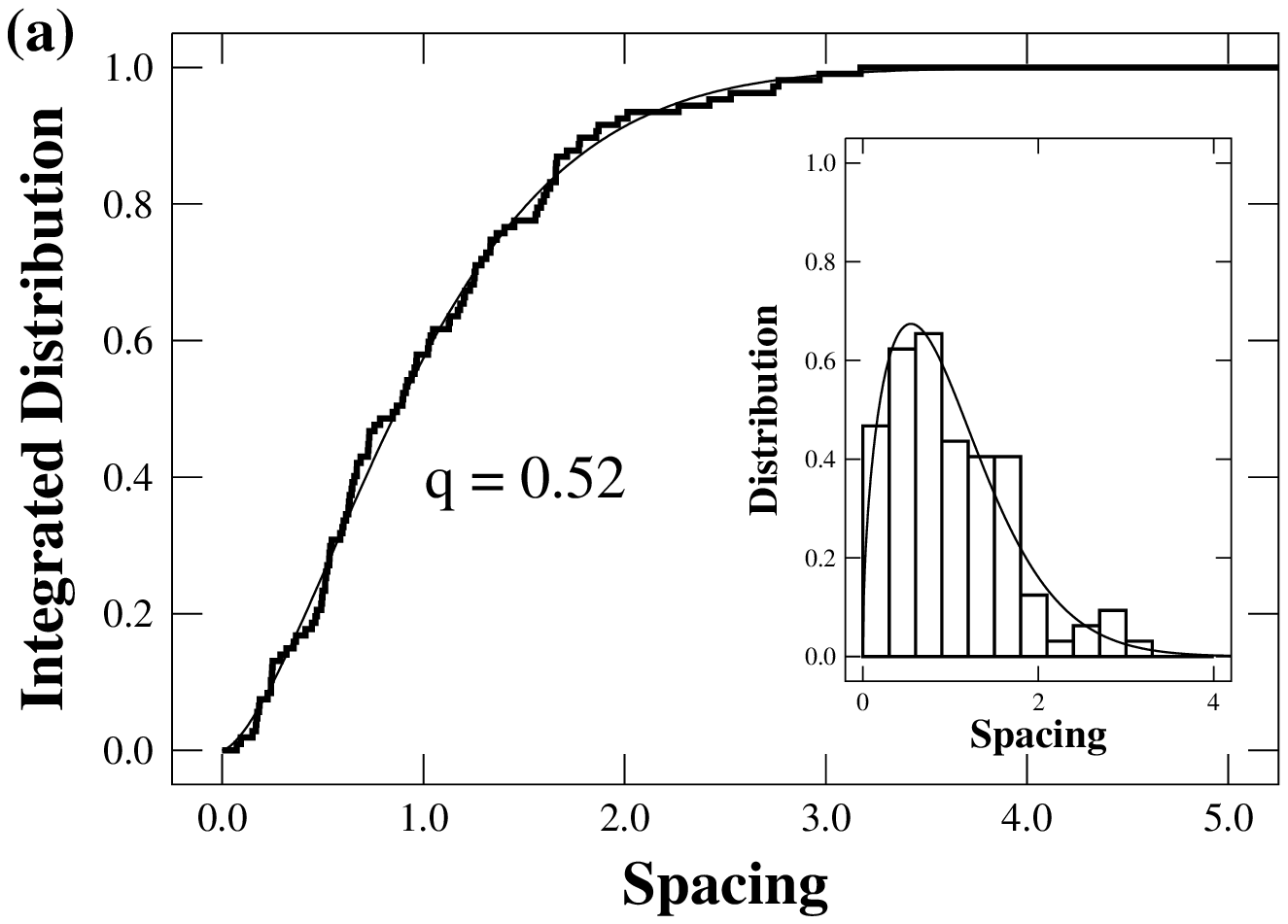}
\includegraphics[scale=0.5]{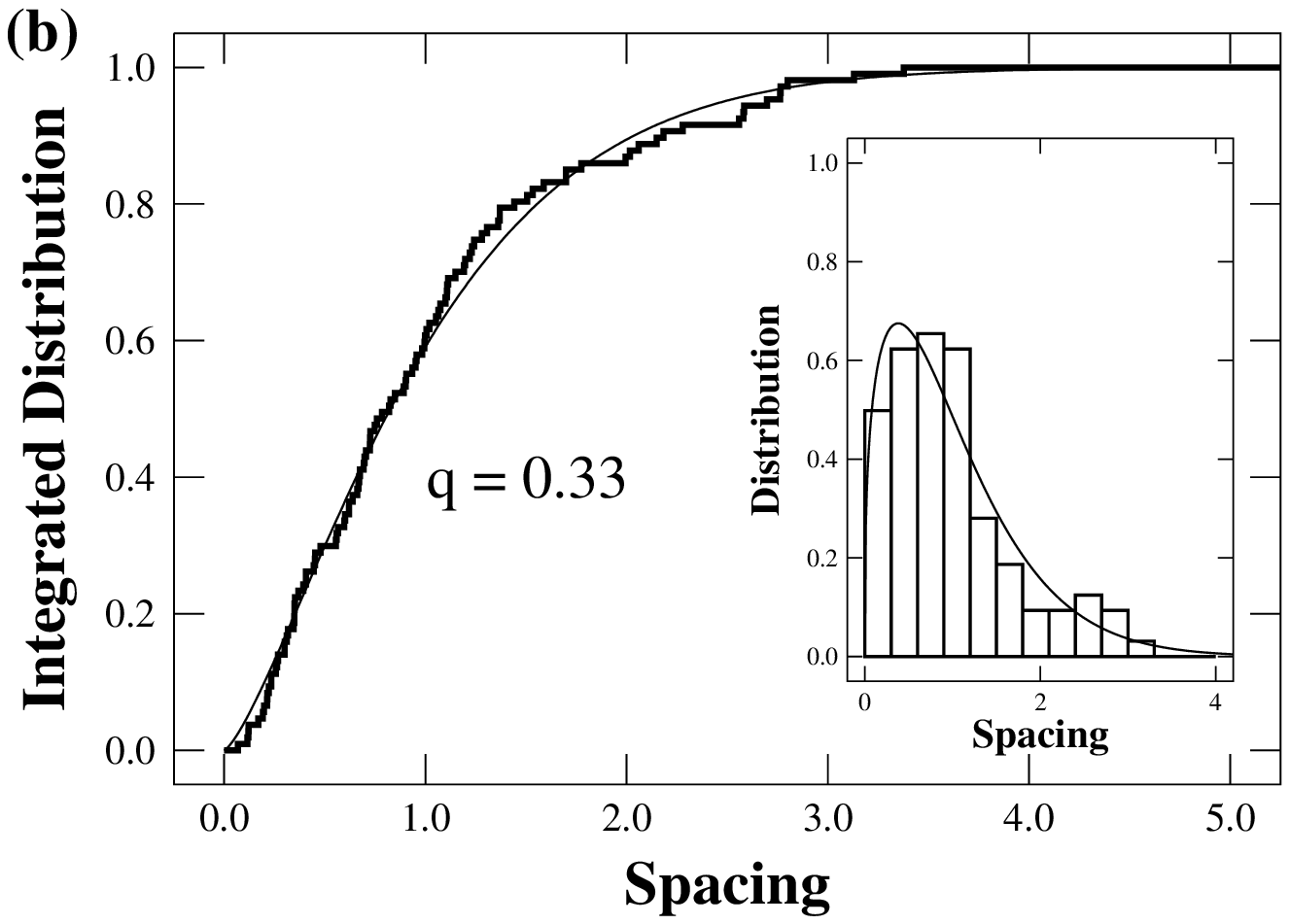}
\caption{The integrated distribution of the nearest-neighbor spacing of
the orbital energies of a met-enkephalin
(a) in vacuum and (b) in water.
We also show histograms for the nearest-neighbor spacing in the insets,
where curves show the Brody distribution (see text)
with (a) $q=0.52$ and (b) $q=0.33$.}
\label{fig:spacing}
\end{center}
\end{figure*}

Despite that these orbital energies calculated from
the generalized eigen-equation (\ref{eqn:roothaan-hall})
are not usual eigenvalues defined in an autonomous system,
they have almost the same property as the usual eigenvalues and eigenstates,
i.e., the orbital energies are real, and the orbital states are
orthogonal to the state with a different orbital energy.
Moreover, repulsive interactions between them
have the same origin as those of the eigenvalues\cite{HF-statistics}.
Thus, we can perform the eigenvalue-statistical analysis
of these values after a certain unfolding procedure.
In Figs. \ref{fig:spacing} (a) and (b),
we show the integrated nearest-neighbor spacing distribution
obtained after the standard procedure of unfolding the spectrum\cite{Haake}.
It is known that the fully chaotic system
shows the Wigner distribution while the regular system shows
the Poisson distribution.
For an intermediate case, we can compare the distribution
to a Brody distribution.
In figs. \ref{fig:spacing} (a) and (b),
we also plot curves for the Brody distribution
which is obtained by fitting the distribution
to the integrated Brody distribution,
\begin{equation}
  P_B(s)=1-\exp(-\alpha s^{q+1})
\end{equation}
where $q$ represents the Brody parameter, and
\begin{equation}
  \alpha\equiv\left[\Gamma\left(\frac{q+2}{q+1}\right)\right]^{q+1}
\end{equation}
is obtained by normalization condition.
The values of the Brody parameters obtained are
$q=0.52$ in vacuum and $q=0.33$ in water.
The smaller value of $q$ reflects the smaller interaction
between orbitals.

In order to study further quantum aspects of the solvation effect,
we enter the detailed analysis of energies and wavefunctions
to the fictitious parameter $\lambda$.
For chaotic quantum systems, the Brody parameter $q$ is
usually related to the area of stochastic region in phase space\cite{Brody}.
However, we do not intend to study any stochastic dynamics
behind the electronic states.
Instead, we concentrate on the interaction between the orbitals
when we slightly vary the parameter $\lambda$,
i.e., avoided crossings\cite{TTavoided} between the orbital energies.

\begin{figure}[b]
\begin{center}
\includegraphics[scale=0.57]{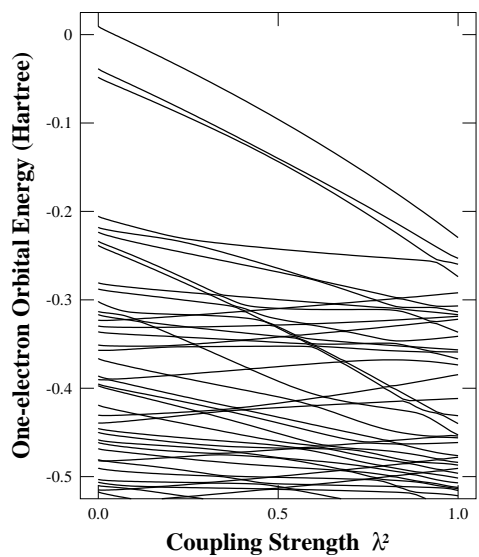}
\caption{One-electron orbital energies near the highest occupied
molecular orbital. The abscissa is $\lambda^2$, the square of
the coupling strength between the solute and solvent.}
\label{fig:avoided}
\end{center}
\end{figure}

The Fock matrix (\ref{eqn:fock}) depends on the fictitious
parameter $\lambda$ nonlinearly
since $\lambda$ is also included in $V_{\mu\nu}$,
which is different from linearly perturbed systems.
Even then, we can plot the orbital energies to the parameter as usual
if we perform SCF calculations from $\lambda=0$ (vacuum)
to $\lambda=1$ (fully solvated).
Figure \ref{fig:avoided} is a plot of the orbital energies
with respect to the parameter $\lambda$,
where we chose $\lambda^2$ instead of $\lambda$ as the abscissa
because of the nonlinear dependence on $\lambda$.
Various sizes of avoided crossing are found in the figure,
and there seems to be a small number of special orbitals
which have almost no interaction with others.
Such a separation of orbital states into groups without interaction
each other can lead to smaller values of the Brody parameter.

The separation of eigenfunctions is often seen in the intermediate
semiclassical limit of quantum chaotic systems,
i.e., statistical properties in a range around a finite energy.
A stadium billiard system has a series of eigenfunctions
corresponding to a bouncing-ball orbit
by which the nearest-neighbor spacing deviates
from the ideal case of random matrix systems.
The eigenstates corresponding to the bouncing-ball orbit
is extremely localized in momentum space\cite{Localize1,Localize2}.
In the present results,
such localization can be observed in the coordinate space.
Actually, the core orbitals, which we have removed from the Brody analysis,
are strongly localized on a certain atomic basis function.
Although the orbitals in the region we study here do not localize so strong,
it is reasonable to analyze the property of orbital states
with respect to the localization on the atomic basis functions.

\begin{figure}[tb]
\begin{center}
\includegraphics[scale=0.5]{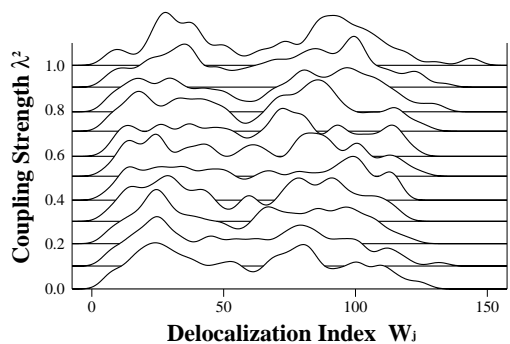}
\caption{Distribution $P(W_j)$ of the delocalization index $W_j$
from 45th to 152nd orbital states.}
\label{fig:delocalize}
\end{center}
\end{figure}

In order to analyze the extent of localization,
we define the degree of delocalization of $j$-th orbital state by
\begin{equation}
\label{eqn:delocalize}
  W_j=\left|\displaystyle\sum_\nu\left|C_{\nu j}\right|\right|^2
        \left/\displaystyle\sum_\nu\left|C_{\nu j}\right|^2\right.
\end{equation}
through the basis-set expansion coefficient $C_{\nu j}$ in (\ref{eqn:basis}).
This is a representation-dependent quantity, and
we define $W_j$ on the standard atom-centered basis
set $\{|\chi_\nu\rangle\}$ in the present work.
Although $\{|\chi_\nu\rangle\}$ is
not orthogonal each other and we cannot relate the coefficients
$C_{\nu j}$ directly to the extent of the spatial distribution,
the value of (\ref{eqn:delocalize}),
the approximate number of basis functions contained,
still represents the relative strength of the delocalization.
In Fig. \ref{fig:delocalize},
we show the distribution of (\ref{eqn:delocalize}).
It can be seen that the distribution has two peaks
over the whole values of $\lambda$
while the places of the peaks slightly change.
That is, the orbitals are divided into two groups:
relatively localized orbitals ($0<W_j<60$) and
delocalized orbitals ($60<W_j<130$).
This separation of orbitals is considered to be the reason why
the Brody parameter exhibits the intermediate values.

Before summarizing this work,
we give some comments for future studies.
Since our analysis is based on the parameter variation\cite{TTslope},
more sophisticated analysis such as the level-curvature analysis
\cite{GRMN1990,TTcurvature,ZD93c} and
distribution of gap sizes of avoided crossings\cite{ZD93a,MS2005}
can be performed.
Since these properties are known to be sensitive for intermediate
dynamical systems,
it will be more appropriate for analyzing molecular systems.
By the use of these analysis,
it is desirable to investigate quantum aspects of large molecules in water.
When we investigate large systems such as proteins or nucleic acids,
more effective methods must be needed.
One of such methods is the fragment MO calculation\cite{FMO}
by which the orbital energies can be obtained with small errors\cite{FMO-MO}.

Another interesting problem is whether the regularity of the
quantum many-body system is influenced by solvation effects or not,
while in the present work
we have concentrated on the localization of molecular orbitals.
Although we could not give a clear answer within our Brody analysis,
such analysis on the regularity will be performed
through the detailed analysis of nodal patterns, Husimi representation,
etc. of the orbital wavefunctions with respect to the fictitious parameter.
Introducing the parameter
enables us to assign each solvated orbital
in terms of the orbitals in vacuum.
This procedure will successfully be done by
connecting characteristic features of the orbital over avoided crossings
since it is possible even in strongly chaotic systems\cite{TTcharacter}
if many pairwise avoided crossings are found.
From a technical point of view,
the continuous parameter is also important as a computational technique.
In our case, the stability of the result is numerically supported
by varying the parameter slightly,
since physically meaningful quantities can only be obtained
as stable results under small perturbations.
In order to obtain stable and significant results
from complicated realistic simulations,
the fictitious parameter approach will be one of the powerful methods.

In summary, we have studied statistics of the orbital energies
and localization of orbital states
through the variation of the fictitious parameter.
This is our first report on the eigenvalue-statistical analysis
of the orbitals as a new approach to quantum many-body systems.
We hope that these new methods will be used to study
quantum aspects of large bio-molecules
as well as to complete large-scale realistic simulations.

\section*{Acknowledgments}

We would like to thank Prof. F. Hirata
for allowing us to use the 3D-RISM program developed in his group.
We also thank Dr. S. Ho, Mr. S. Kubo and Mr. K. Fukuda for
their collaborative development on the 3D-RISM/SCF coupled simulation.
One of the authors (T.T.) is grateful to Prof. S. Sawada, Prof. M. Toda,
Prof. A. Shudo, Dr. M. Sano and Dr. A. Tanaka
for extensive discussion on chaos and dynamical systems.
This work is supported by the Ministry of Education, Sports, Culture,
Science and Technology (MEXT) through the Science-grid NAREGI Program
under the Development and Application of Advanced High-performance
Supercomputer Project.

\end{document}